\documentstyle[12pt]{article}

\newcommand{\beq}{\begin{equation}}
\newcommand{\eeq}{\end{equation}}
\newcommand{\bea}{\begin{eqnarray}}
\newcommand{\ena}{\end{eqnarray}}
\newcommand{\OMP}{\mbox{$\omega_{0}$}}
\newcommand{\OMT}{\mbox{$\omega_{T}$}}
\newcommand{\OML}{\mbox{$\omega_{L}$}}

\begin{document}
\begin{titlepage}
\begin{center}
\rightline{CERN-TH-7076/93}
\rightline{ENSLAPP-A-444/93}

\bigskip
\bigskip
\bigskip
\bigskip

{\LARGE \bf {The Electric Charge of Neutrinos and Plasmon Decay}}\\
\vskip 0.8cm
{\large T.~Altherr \footnote{On leave of absence from L.A.P.P.,
BP~110, F-74941, Annecy-le-Vieux Cedex, France}}\\
{\em Theory Division, CERN, CH-1211 Geneva 23, Switzerland}\\
\medskip
{\em and}\\
\medskip
{\large P.~Salati}\\
{\em ENSLAPP, BP110, F-74941 Annecy-le-Vieux Cedex, France}\\
{\em Institut universitaire de France}\\
\end{center}
\bigskip

\begin{abstract}
By using both thermal field theory and a somewhat more intuitive method,
we define the electric charge as well as the charge radius of neutrinos
propagating inside a plasma. We show that electron neutrinos acquire a
charge radius of order $\sim 6.5 \times 10^{-16}$ cm, regardless of the
properties of the medium. Then, we compute the rate of plasmon decay
which such an electric charge or a charge radius implies. Taking into
account the relativistic effects of the degenerate electron gas, we compare
our results to various approximations as well as to recent
calculations  and determine the regimes where the electric charge or the
charge radius does mediate the decay of plasmons. Finally, we discuss the
stellar limits on any anomalous charge radius of neutrinos.
\end{abstract}
\vskip 2.cm
\leftline{CERN-TH-7076/93}
\leftline{November 93}
\end{titlepage}

\section{Introduction}

The plasmon decay process is an important cooling mechanism for
stars whose core is degenerate. It is particularly relevant as regards the
evolution of young white dwarves and red giant stars. In these two stellar
objects, the core is composed of degenerate matter, with density
$\rho\simeq 10^6$ g cm$^{- 3}$ and temperature in the range from $T =
10^{7}$ to $10^{9}$ K. As a consequence of this degeneracy, all electronic
levels below the Fermi sphere are filled up. A typical value of the electron
momentum is $p_{F} \sim 400$ keV, so that electrons are neither
non-relativistic, nor ultra-relativistic.

The plasmon decay process was first considered a long time ago, in the
pioneering work by Adams, Ruderman and Woo \cite{ARW}, which led to
a rich literature \cite{Plasmon}. The subject has recently been revived when
it was realized that the plasmon dispersion relations, which were used
previously, did not incorporate the relativistic behaviour of the electron
plasma \cite{Bra}. The latest works that deal with these effects are due to
Itoh {\it et al.} \cite{IMHK}, Braaten and Segel \cite{BS} and Haft, Raffelt
and Weiss \cite{HRW}.

In this paper, we complete the previous analysis and present the calculation
of the plasmon decay rate in the framework of two different but
complementary approaches. The first discussion is based on considerations
of thermal field theory while the second analysis is a more intuitive, kinetic
theory-like, method. We show that both reasonings lead to the same
answer. We also stress the important distinction between the effective
electric charge which neutrinos exhibit when they couple to a classical
electromagnetic field and their charge radius which may be strictly defined
in the soft regime where the energy and momentum of plasmons are small.

The content of this work is as follows. In section~2, we derive the electric
charge which a neutrino acquires when it propagates inside matter and
interacts with a classical electromagnetic field. In section~3, we discuss the
charge radius associated to the coupling between neutrinos and low-energy
plasmons. We comment on these two phenomenological approximations
which lead, a priori, to fairly different descriptions of the electromagnetic
properties of neutrinos. In section~4, we present the analytic expressions for
the plasmon decay process. We discuss our numerical results for the case of
red-giants and white dwarves in section~5 where various approximations
to the energy loss rate are presented along with their respective domain of
validity. We finally conclude that, for most astrophysical purposes, the
charge radius and the electric charge descriptions of the neutrino behaviour,
give indeed the same answer.

\section{The neutrino electric charge and the hard regime}

When neutrinos propagate inside matter, they scatter coherently on charged
particles, such as the electrons and positrons of the thermal bath. This
process is responsible for the ``additional'' inertia which neutrinos possess
in a plasma \cite{RN}. Another possibility is that neutrinos diffuse
coherently on electrons and positrons which, in turn, couple to the
electromagnetic field $A^{\beta}$ \cite{Charge}. The Feynman graphs
responsible for the latter reaction are presented in fig.~1 where both
electrons (fig.~1a and 1b) and positrons (fig.~1c and 1d) have been taken into
account. Note that the final electronic (or positronic) state is just the same
 as
the incoming one, hence a modification of the electromagnetic coupling of
neutrinos which appears directly at the level of the amplitude of this
coherent diffusion. The effective coupling of neutrinos to electrons results
from the low-energy limit of the Standard Glashow-Weinberg-Salam
theory~:
\beq
{\cal L} \; = \; \frac{G_{F}}{\sqrt{2}} \;\;
\left\{ \bar{\Psi}_{\nu} \gamma_{\alpha} \left( 1 - \gamma_{5} \right)
\Psi_{\nu} \right\} \;\;
\left\{ \bar{\Psi}_{e} \gamma^{\alpha}
\left( g_{V} - g_{A} \gamma_{5} \right) \Psi_{e} \right\} \;\; .
\eeq
Both $Z$ and $W$ exchanges are taken into account for electron neutrinos
so that
\beq
g_{V} \; = \; \frac{1}{2} \, + \, 2 \, \sin^{2} \theta_{W}
\;\;\;\; {\rm and} \;\;\;\;
g_{A} \; = \; \frac{1}{2} \;\; ,
\eeq
whilst, for $\nu_{\mu}$ and $\nu_{\tau}$ where the $Z$ boson alone
mediates the interaction, the coupling parameters are
\beq
g_{V} \; = \; - \frac{1}{2} \, + \, 2 \, \sin^{2} \theta_{W}
\;\;\;\; {\rm and} \;\;\;\;
g_{A} \; = \; - \frac{1}{2} \;\; .
\eeq
In order to compute the effective coupling of neutrinos to the
electromagnetic field $A^{\beta}$ inside a plasma, the spectator electrons
must be integrated out. Let us consider a volume $V$ of matter. The
electronic state $\Psi_{e}$ describes the presence of a single electron of
momentum $\vec{p \,}$ in the whole volume $V$, hence the
normalisation
\beq
\Psi_{e}(\vec{p \,}, s) \; = \;
\sqrt{\frac{m}{VE}} \; u(\vec{p \,}, s) \;\; ,
\eeq
where $m$ and $E$ stand respectively for the mass and the energy of the
electron. For positrons, $u(\vec{p \,}, s)$ is replaced by $v(\vec{p \,}, s)$.
The
electromagnetic coupling of neutrinos may therefore be expressed as a sum
of the amplitude of the coherent diffusion pictured in fig.~1, over the
electronic and positronic states of the thermal bath~:
\bea
{\cal L}_{eff} & = & \frac{G_{F}}{\sqrt{2}} \;\;
\left\{ \bar{\Psi}_{\nu} \gamma_{\alpha} \left( 1 - \gamma_{5} \right)
\Psi_{\nu} \right\} \;\; A_{\beta} \;\; {\displaystyle \int} \;
\frac{V \, d^{3} \vec{p \,}}{8 \pi^{3}} \; {\displaystyle \sum}_{s} \;
\times \label{A} \\
& \eta_{e}(\vec{p \,}) & \left\{
\begin{array}{c}
\left[ \bar{\Psi}_{e}(\vec{p \,} , s) \gamma^{\alpha}
\left( g_{V} - g_{A} \gamma_{5} \right)
{\displaystyle \frac{1}{({\rlap/P} + {\rlap/Q} - m)}} e \gamma^{\beta}
\Psi_{e}(\vec{p \,} , s) \right] \\
+ \; \left[ \bar{\Psi}_{e}(\vec{p \,} , s) e \gamma^{\beta}
{\displaystyle \frac{1}{({\rlap/P} - {\rlap/Q} - m)}} \gamma^{\alpha}
\left( g_{V} - g_{A} \gamma_{5} \right) \Psi_{e}(\vec{p \,} , s) \right]
\end{array}
\right\} \nonumber \\
- & \eta_{\bar{e}}(\vec{p \,}) & \; \left\{
\begin{array}{c}
\left[ \bar{\Psi}_{\bar{e}}(\vec{p \,} , s) e \gamma^{\beta}
{\displaystyle \frac{1}{(- {\rlap/P} - {\rlap/Q} - m)}} \gamma^{\alpha}
\left( g_{V} - g_{A} \gamma_{5} \right) \Psi_{\bar{e}}(\vec{p \,} , s)
\right] \\
+ \; \left[ \bar{\Psi}_{\bar{e}}(\vec{p \,} , s) \gamma^{\alpha}
\left( g_{V} - g_{A} \gamma_{5} \right)
{\displaystyle \frac{1}{(- {\rlap/P} + {\rlap/Q} - m)}} e \gamma^{\beta}
\Psi_{\bar{e}}(\vec{p \,} , s) \right]
\end{array}
\right\} \nonumber \;\; .
\ena
Here, $P^{\mu} = (E , \vec{p \,})$ denotes the electron (or positron)
momentum whilst $Q^{\mu} = (\omega , \vec{q \,})$ refers to the
plasmon.
Since the initial and final states of the spectator electrons are the same,
expression (\ref{A}) may be written as a sum of traces. The statistical
occupation numbers of electrons and positrons are respectively taken care of
by the Fermi-Dirac functions $\eta_{e}(\vec{p \,})$ and
$\eta_{\bar{e}}(\vec{p \,})$. The effective Lagrangian (\ref{A}) becomes
\bea
{\cal L}_{eff} & = & \frac{e \, G_{F}}{\sqrt{2}} \;\;
\left\{ \bar{\Psi}_{\nu} \gamma_{\alpha} \left( 1 - \gamma_{5} \right)
\Psi_{\nu} \right\} \;\; A_{\beta} \;\; {\displaystyle \int} \; \tilde{dp} \;
\times \\
& \eta_{e}(\vec{p \,}) & \left\{
\begin{array}{cl}
& {\displaystyle \frac{{\rm Tr} \left[ \left( {\rlap/P} + m \right)
\gamma^{\alpha} \left( g_{V} - g_{A} \gamma_{5} \right)
\left( {\rlap/P} + {\rlap/Q} + m \right) \gamma^{\beta} \right]}
{Q^{2} \; + \; 2 \, P \! \cdot \! Q}} \\
+ & \\
& {\displaystyle \frac{{\rm Tr} \left[ \left( {\rlap/P} + m \right)
\gamma^{\beta} \left( {\rlap/P} - {\rlap/Q} + m \right) \gamma^{\alpha}
\left( g_{V} - g_{A} \gamma_{5} \right) \right]}
{Q^{2} \; - \; 2 \, P \! \cdot \! Q}}
\end{array}
\right\} \nonumber \\
- & \eta_{\bar{e}}(\vec{p \,}) & \left\{
\begin{array}{cl}
& {\displaystyle \frac{{\rm Tr} \left[ \left( {\rlap/P} - m \right)
\gamma^{\beta} \left( - {\rlap/P} - {\rlap/Q} + m \right)
\gamma^{\alpha} \left( g_{V} - g_{A} \gamma_{5} \right) \right]}
{Q^{2} \; + \; 2 \, P \! \cdot \! Q}} \\
+ & \\
& {\displaystyle \frac{{\rm Tr} \left[ \left( {\rlap/P} - m \right)
\gamma^{\alpha} \left( g_{V} - g_{A} \gamma_{5} \right)
\left( - {\rlap/P} + {\rlap/Q} + m \right) \gamma^{\beta} \right]}
{Q^{2} \; - \; 2 \, P \! \cdot \! Q}}
\end{array}
\right\} \nonumber \;\; ,
\ena
where $\tilde{dp}$ is the Lorentz-invariant differential element
\beq
\tilde{dp} \; = \; \frac{1}{2E} \, \frac{d^{3} \vec{p \,}}{8 \pi^{3}} \;\; .
\eeq
After the reduction of the above-mentionned traces and some
straightforward algebra, the effective coupling of neutrinos to the
electromagnetic field may be expressed as
\beq
{\cal L}_{eff} \; = \; \frac{G_{F}}{\sqrt{2}} \;\;
\left\{ \bar{\Psi}_{\nu} \gamma_{\alpha} \left( 1 - \gamma_{5} \right)
\Psi_{\nu} \right\} \;\; \Gamma^{\alpha \beta} \;\; A_{\beta} \;\; ,
\label{B}
\eeq
where the tensor $\Gamma^{\alpha \beta}$ may be separated into its
symmetric and antisymmetric components~:
\bea
\Gamma^{\alpha \beta} & = & 4 e g_{V} \, {\displaystyle \int}
\tilde{dp} \left\{ \eta_{e}(\vec{p \,}) + \eta_{\bar{e}}(\vec{p \,})
\right\} \,
\left\{{\displaystyle \frac{\left( P \! \cdot \! Q \right)
\left(P^{\alpha} Q^{\beta} + P^{\beta} Q^{\alpha} \right) -
\left( P \! \cdot \! Q \right)^{2} g^{\alpha \beta} -
Q^{2} P^{\alpha} P^{\beta}}
{\left( P \! \cdot \! Q \right)^{2} \; - \; \left( Q^{2} / 2 \right)^{2}}}
 \right\}
\nonumber \\
& - & 2 i e g_{A} \, \epsilon^{\alpha \beta \mu \nu} \,
{\displaystyle \int} \tilde{dp}
\left\{ \eta_{e}(\vec{p \,}) \, - \, \eta_{\bar{e}}(\vec{p \,}) \right\} \;
\left\{{\displaystyle \frac{P_{\mu} Q_{\nu} Q^{2}}
{\left( P \! \cdot \! Q \right)^{2} \; - \; \left( Q^{2} / 2 \right)^{2}}}
 \right\}
\;\; .
\label{C}
\ena
This effective coupling $\Gamma^{\alpha \beta}$ has been derived fairly
intuitively. As pictured in fig.~1, the neutrino wave interacts coherently
with the charged species of the plasma -- for that matter mostly electrons
and positrons since the contribution of nuclei is negligible. Expression
(\ref{C}) may be directly obtained from mere considerations of thermal field
theory.

At the one-loop level, the Feynman rules for finite temperature field
theory are quite simple \cite{LvW}. The only ingredient is here the electron
propagator, given by
\beq
S_{F}(P) \; = \; \left( {\rlap/P} + m \right) \;
\left[ \frac{i}{P^2 - m^2 + i\epsilon} \; - \;
2 \pi \delta(P^2 - m^2) N_{F}(P) \right] \;\; ,
\eeq
where the statistical factor $N_{F}(P)$ may be expressed as
\beq
N_{F}(P) \; = \;
\frac{ \theta(P_{0}) }{ e^{\displaystyle (P_{0} - \mu)/T} + 1 }
\; + \;
\frac{ \theta(- P_{0}) }{ e^{\displaystyle (- P_{0} + \mu)/T} + 1 } \;\; .
\eeq
Using the same conventions as before, the only non-vanishing contribution
to
the one-loop diagram of fig.~2 is
\bea
\Gamma^{\alpha \beta} & = & e {\displaystyle \int} \;
\frac{d^4 P}{(2 \pi)^3} \;
{\rm Tr} \left[ \gamma^{\alpha} \left( g_{V} - g_{A} \gamma_{5} \right)
\left( {\rlap/P} + m \right) \gamma^{\beta}
\left( {\rlap/P} + {\rlap/Q} + m \right) \right] \nonumber \\
& \times & \left\{
\frac{\delta(P^2 - m^2) N_{F}(P)}{(P+Q)^2-m^2} \; + \;
\frac{\delta[(P+Q)^2 - m^2] N_{F}(P+Q)}{P^2-m^2} \right\} \;\; ,
\ena
which, after a change of variables, reduces to relation (\ref{C}). The same
result could also have been derived in the imaginary-time formalism.

The effective electric charge of the neutrinos which propagate inside a
plasma may be defined as the limit of expression (\ref{C}) where
$A^{\beta}$ describes a classical electromagnetic field. The behaviour of
finite temperature field theory and especially of its perturbative approach of
interactions is quite rich. Inside matter, photons no longer propagate at the
speed of light. On the contrary, they follow dispersion relations which imply
the existence of longitudinal and transverse modes -- see equations
(\ref{LONGITUDINAL}) and (\ref{TRANSVERSE}). The transverse
photons
are extremely similar to the ordinary quanta of the vacuum. They should be
disentangled from the longitudinal modes which are mere collective
excitations of the plasma. That is why the transverse branch alone exhibits
an asymptote where the energy $\omega$ and the momentum $\vec{q \,}$
of
plasmons reach up to infinity at constant $Q^{2} = \omega^{2} - q^{2}$. In
this so-called {\it hard limit} \cite{HARDSOFT}, the transverse photons
behave as if they propagated in the vacuum, with the non-vanishing mass
$m_{T}$. Note that $Q^{2}$ is always of order $\omega_{0}^{2}$, where the
plasma frequency $\omega_{0}$ is smaller than the electron mean energy
$<E_{e}>$ by a factor of $e = \sqrt{4 \pi \alpha}$, {\it i.e.}, $\omega_{0}
\sim e <E_{e}> \sim e \, T$ or $e \, p_{F}$. Therefore, in the hard regime,
the components $\omega$ and $q$ may vary regardless of $Q^{2}$ which
behaves as a small perturbation of order $\alpha$. In relation (\ref{C}),
expressions such as $(P \! \cdot \! Q)$ or $P^{\alpha} Q^{\beta}$ may be
considered as leading contributions with respect to the correction $Q^{2}
\sim O(\alpha)$. The tensor $\Gamma^{\alpha \beta}$ may therefore be
expanded as a function of the small parameter $Q^{2}$, and the leading term
turns out to be the symmetric expression
\bea
\Gamma^{\alpha \beta} & \simeq & 4 e g_{V} \, {\displaystyle \int}
\tilde{dp} \left\{ \eta_{e}(\vec{p \,}) \, + \, \eta_{\bar{e}}(\vec{p \,})
\right\} \; \left\{ {\displaystyle
\frac{P^{\alpha} Q^{\beta} + P^{\beta} Q^{\alpha} -
\left( P \! \cdot \! Q \right) g^{\alpha \beta}}{P \! \cdot \! Q}} \right\}
\nonumber \\
& + & {\rm terms \; of \; order} \; \alpha \;\; .
\label{D}
\ena
As expression (\ref{D}) is folded into the vertex (\ref{B}), it further
simplifies. In the Lorentz gauge where $\partial_{\beta} A^{\beta} = 0$, the
contribution $Q^{\beta} A_{\beta}$ vanishes. Note that the effective
interaction (\ref{B}) is gauge invariant since $\Gamma^{\alpha \beta}
Q_{\alpha} = \Gamma^{\alpha \beta} Q_{\beta} = 0$, as is obvious from
expression (\ref{C}). Therefore, restricting ourselves to the Lorentz gauge
does not alter the general import of our reasoning. Finally, neutrinos may
be considered as massless species, a fairly good approximation since
plasmon decay merely involves the electron neutrino for which $g_{V}
\simeq 1$ and whose mass is orders of magnitude smaller than the typical
energies at stellar cores. Since expression $\bar{\Psi}_{\nu} {\rlap/Q} \left(
1 - \gamma_{5} \right) \Psi_{\nu}$ vanishes in that limit, the neutrino
coupling further simplifies into the effective Lagrangian~:
\beq
{\cal L}_{eff} \; = \; - \; e Q_{\nu} \;
\left\{ \bar{\Psi}_{\nu} \gamma^{\beta} L \Psi_{\nu} \right\} \;
A_{\beta} \;\; .
\label{E}
\eeq
The left-helicity operator $L$ is defined as $L = \left( 1 - \gamma_{5} \right)
/ 2$. The neutrino behaves as if it had the electric charge
\beq
Q_{\nu} \; = \; 4 \sqrt{2} \, G_{F} g_{V} {\cal I}_{P} \;\; .
\label{QNU}
\eeq
This charge depends on the properties of the plasma through the integral
\beq
{\cal I}_{P} \; = \; \frac{1}{4 \pi^{2}} \; {\displaystyle \int} \;
\frac{p^{2} dp}{E} \; \left\{ \eta_{e}(p) \, + \, \eta_{\bar{e}}(p) \right\}
\;\; .
\eeq
In a degenerate gas of electrons, a typical situation inside red-giants or
white
dwarves, this plasma integral over the statistical distribution of electrons
may be expressed as
\beq
{\cal I}_{P} \; = \; \frac{p_{F} E_{F}}{8 \pi^{2}} \;
\left\{ 1 \; + \; \left( \frac{v_{F}^{2} - 1}{2 v_{F}} \right) \,
\ln \left( \frac{1 + v_{F}}{1 - v_{F}} \right) \right\} \;\; ,
\label{IPLASMA}
\eeq
where $v_{F} = p_{F} / E_{F}$ denotes the velocity at the surface of the
Fermi sea. At the helium core of red-giant stars, for instance, the density
reaches up to $\sim 10^{6} \; {\rm g \, cm^{-3}}$. The Fermi momentum is
$p_{F} \sim$ 410 keV, hence an effective electric charge $Q_{\nu} \sim 6.4
\times 10^{-14}$ which, in spite of being some 13 orders of magnitude
smaller than the charge of the electron, induces a significant energy loss, so
large that it dominates the evolution of stars during their ascension of the
red-giant branch.

\section{The neutrino charge radius and the soft limit}

In the low energy part of the transverse branch as well as all along the
longitudinal branch, the energy $\omega$ and the momentum $q$ of
plasmons are much smaller than the average energy of the charged species
of the thermal bath. In this so-called {\it soft regime} \cite{HARDSOFT}
where $q$ is small, the components $Q^{\mu}$ of the plasmon
four-momentum may be considered of order $e = \sqrt{4 \pi \alpha}$ with
respect to the typical electron momentum, {\it i.e.},
\beq
\frac{\left| Q^{\mu} \right|}{\left| P^{\nu} \right|} \; \sim \;
{\cal O}(e) \; \sim \; {\cal O}(\sqrt{4 \pi \alpha}) \; \ll \; 1 \;\; .
\eeq
Notice that in the hard regime, $Q^{2}$ is of order ${\cal O}(\alpha)$
whereas both components $Q^{\mu}$ and $P^{\nu}$ are leading order
terms. On the contrary, in the soft regime which we discuss now,
expressions such as $\omega$, $q$ or $\sqrt{\omega^{2} - q^{2}}$ are of
order ${\cal O}(e)$, so that the perturbative expansion of expression
(\ref{C}) is now quite different than in the hard-loop approximation. In
particular, the axial part of the tensor $\Gamma^{\alpha \beta}$ is of order
${\cal O}(e)$. However, it does not interfere with the vector component
of $\Gamma^{\alpha \beta}$ as regards plasmon decay, and therefore
contributes a term only of order ${\cal O}(\alpha)$ which we will not
consider in what follows. The leading contribution to $\Gamma^{\alpha
\beta}$ is the vector component~:
\bea
\Gamma^{\alpha \beta} & = & 4 e g_{V} \, {\displaystyle \int}
\tilde{dp} \left\{ \eta_{e}(\vec{p \,}) + \eta_{\bar{e}}(\vec{p \,})
\right\} \,
\left\{{\displaystyle \frac{\left( P \! \cdot \! Q \right)
\left(P^{\alpha} Q^{\beta} + P^{\beta} Q^{\alpha} \right) -
\left( P \! \cdot \! Q \right)^{2} g^{\alpha \beta} -
Q^{2} P^{\alpha} P^{\beta}}
{\left( P \! \cdot \! Q \right)^{2}}} \right\} \nonumber \\
& + & {\rm terms \; of \; order} \; e\;\; .
\label{SOFTVERTEX}
\ena
As noticed already some thirty years ago \cite{ARW}, the previous
expression is merely the polarization tensor $\Pi^{\alpha \beta}$ associated
to the propagation of electromagnetic waves inside a plasma, {\it i.e.},
\beq
\Gamma^{\alpha \beta} \; = \; \frac{g_{V}}{e} \, \Pi^{\alpha \beta} \;\; ,
\eeq
hence the effective Lagrangian
\beq
{\cal L}_{eff} \; = \; \frac{\sqrt{2} G_{F} g_{V}}{e} \;
\left\{ \bar{\Psi}_{\nu} \gamma_{\alpha} L \Psi_{\nu} \right\} \;
\Pi^{\alpha \beta} \; A_{\beta} \;\; ,
\label{F}
\eeq
where the electromagnetic field couples to the neutrino current through the
polarization tensor. Note that $\Pi^{\alpha \beta}$ may be broken up into
the operators $P_{T}^{\alpha \beta}$ and $P_{L}^{\alpha \beta}$ which
respectively project the polarization $A^{\beta}$ on the transverse and
longitudinal modes \cite{Wel}~:
\beq
\Pi^{\alpha \beta} \; = \;
\Pi_{T}(\omega, \vec{q \,}) \, P_{T}^{\alpha \beta} \; + \;
\Pi_{L}(\omega, \vec{q \,}) \, P_{L}^{\alpha \beta} \;\; .
\eeq
Since the plasmon four-momentum follows the dispersion relation
\beq
Q^{2} \, + \, \Pi_{T, \, L}(\omega, \vec{q \,}) \; = \; 0 \;\; ,
\label{DISPERSION}
\eeq
expression (\ref{F}) simplifies further into~:
\beq
{\cal L}_{eff} \; = \; - \, \frac{\sqrt{2} G_{F} g_{V}}{e} \; Q^{2} \;
\left\{ \bar{\Psi}_{\nu} \gamma^{\beta} L \Psi_{\nu} \right\} \;
A_{\beta} \;\; .
\label{G}
\eeq
Therefore, in the soft regime, neutrinos interact with the electromagnetic
field as if they had a non-vanishing charge radius~:
\beq
{\cal L}_{eff} \; = \; - \, e \; \frac{<r^{2}>}{6} \; Q^{2} \;
\left\{ \bar{\Psi}_{\nu} \gamma^{\beta} L \Psi_{\nu} \right\} \;
A_{\beta} \;\; ,
\label{CRC}
\eeq
with the plasmon four-momentum $Q^{2}$ factored out in the coupling, as
is typical. This thermal charge radius, which neutrinos built from their
coherent diffusion on the spectator electrons of the plasma, may be
expressed as
\beq
<r^{2}> \; = \; \frac{3 G_{F}}{\sqrt{2} \pi \alpha} \; g_{V} \;\; ,
\eeq
and may be understood as the superimposition of two distributions of
opposite electric charge, with same centers, but with different spatial
extensions. By surfing on the electrons of the plasma, the neutrino behaves
as if it acquired an inner structure whose typical size is surprisingly
independent of the properties of the thermal bath~:
\beq
<r^{2}> \; \simeq \; \left(6.5 \times 10^{-16} \; {\rm cm}\right)^{2} \;
g_{V} \;\; .
\eeq
We therefore conclude that, in the soft limit, neutrinos interact with the
electromagnetic field through a charge radius coupling whereas, in the hard
regime, they possess an effective electric charge.

In a degenerate medium, the plasma frequency $\omega_{0}$ is related to
the Fermi momentum $p_{F}$
\beq
\omega_{0}^{2} \; = \; \frac{4 \alpha}{3 \pi} \; \frac{p_{F}^{3}}{E_{F}} \;\; .
\label{OMEGAPLASMA}
\eeq
A fully relativistic treatment of the electron bath leads to the dispersion
relation for the longitudinal mode \cite{BS,APR}~:
\beq
\omega_{L}^{2} \; = \; \omega_{0}^{2} \, \left(
\frac{3 \omega_{L}^{2}}{v_{F}^{2} q^{2}} \right) \, \left\{
\left( \frac{\omega_{L}}{2 \, v_{F} q} \right) \,
\ln \left( \frac{\omega_{L} + v_{F} q}{\omega_{L} - v_{F} q} \right)
\, - \, 1 \right\} \;\; ,
\label{LONGITUDINAL}
\eeq
where $v_{F} = p_{F} / E_{F}$ stands for the velocity at the surface of the
Fermi sea. The energy and the momentum of plasmons are respectively
denoted by $\omega_{L}$ and $q$. The longitudinal branch extends up to
the point where it crosses the light cone, for
\beq
q_{\, max} \; = \; \left\{ \frac{3}{v_{F}^{2}} \left( \frac{1}{2 v_{F}} \,
\ln \left[ \frac{1 + v_{F}}{1 - v_{F}} \right] \; - \; 1 \right) \right\}^{1/2}
 \;
\omega_{0} \;\; .
\eeq
As $q$ increases from 0 to $q_{\, max}$, the four-momentum $Q^{2}$
steadily decreases from the plasma frequency $\OMP^{2}$ down to 0. Note
that $\omega_{L}$ and $q$ cannot reach up to infinity, except in the
ultra-relativistic case where, nevertheless, they cannot vary regardless of the
value of $Q^{2}$. The longitudinal branch corresponds therefore to the
soft regime. In particular, when electrons are non-relativistic, {\it i.e.},
for
vanishing $v_{F}$, relation (\ref{LONGITUDINAL}) simplifies into
\beq
\OML = \OMP \;\; ,
\label{LONGITUDINALNR}
\eeq
and the longitudinal branch extends up to $q_{\, max} = \OMP$.
The non-relativistic regime is a good example of soft behaviour in so far as
$\OML$ and $q$, which cannot exceed $\OMP$, are obviously of order
${\cal O}(e)$ with respect to the Fermi momentum $p_{F}$. As regards the
transverse mode, the dispersion relation inside degenerate matter may be
expressed directly as a function of the effective plasmon mass~:
\beq
\omega_{T}^{2} - q^{2} \; = \; \omega_{0}^{2} \,
\left( \frac{3 \omega_{T}^{2}}{2 v_{F}^{2} q^{2}} \right) \, \left\{ 1 \, + \,
\left( \frac{v_{F}^{2} q^{2}}{\omega_{T}^{2}} \, - \, 1 \right)
\left( \frac{\omega_{T}}{2 v_{F} q} \right)
\ln \left( \frac{\omega_{T} + v_{F} q}{\omega_{T} - v_{F} q} \right)
\right\} \;\; .
\label{TRANSVERSE}
\eeq
The low-energy part of the transverse branch corresponds to the soft regime,
whereas the hard limit is recovered for $\omega_{T} \simeq q \to +
\infty$. When the hard regime obtains, transverse plasmons behave as if
they propagated with the effective mass
\beq
m_{T}^{2} \; = \; \left( \frac{3 \omega_{0}^{2}}{2 v_{F}^{2}} \right) \,
\left\{ 1 \, + \,
\left( \frac{v_{F}^{2} \, - \, 1}{2 v_{F}} \right)
\ln \left( \frac{1 + v_{F}}{1 - v_{F}} \right) \right\} \;\; .
\label{MT}
\eeq
As $q$ increases along the transverse branch, note that the four-momentum
$Q^{2}$ raises from $\OMP^{2}$ to $m_{T}^{2}$.
As is clear from relations (\ref{IPLASMA}), (\ref{OMEGAPLASMA}) and
(\ref{MT}), this asymptotic mass may be readily expressed as a function of
the integral ${\cal I}_{P}$ which dominates the behaviour of the neutrino
electric charge $Q_{\nu}$~:
\beq
m_{T}^{2} \; = \; 4 \, e^{2} \, {\cal I}_{P} \;\; .
\eeq
When the stellar plasma is ultra-relativistic, the transverse mass of
high-energy ({\it i.e.}, hard) plasmons is $m_{T} = \sqrt{3/2} \,
\OMP$. On the contrary, for non-relativistic electrons, relation
(\ref{TRANSVERSE}) becomes
\beq
\OMT^{2} = \OMP^{2} + q^{2} \;\; ,
\label{TRANSVERSENR}
\eeq
and plasmons propagate with the mass $\OMP$ all along the transverse
branch, even in its soft region.

Both longitudinal and transverse dispersion
relations were first derived by Jancovici \cite{Jan}, but had a much more
complicated form. It was realized later that the simplified expressions
(\ref{LONGITUDINAL}) and (\ref{TRANSVERSE}) could be used in a
much wider regime \cite{BS,APR}.

A thorough inspection of relation (\ref{C}) shows that equation (\ref{D})
behaves as the mere limit of equation (\ref{SOFTVERTEX}) for vanishing
$Q^{2}$, at fixed plasmon momentum. The transition between the soft and
hard behaviours is therefore smooth. The expressions derived in this
section are surprisingly more general than what could be naively guessed.
Not only are they valid in the soft limit, but they cover the entire plasmon
spectrum and, as a consequence, the hard regime as well.
The effective charge
radius is a good description of the electromagnetic properties of neutrinos.
To illustrate this point, let us consider the effective interaction (\ref{G})
which has been strictly derived in the soft regime. For high energy
plasmons of the transverse mode, $Q^{2}$ may be replaced by $4 \, e^{2} \,
{\cal I}_{P}$ so that the charge radius coupling $(<r^{2}> Q^{2} / 6)$
translates into the electric charge $Q_{\nu}$ of relation (\ref{QNU}). The
charge radius, which has been defined in the soft regime, pleasantly behaves
as an effective electric charge when the hard limit is taken. More generally,
it translates, along the asymptote of the transverse branch, into the electric
charge
\beq
Q_{\nu} \; = \; \frac{G_{F} g_{V}}{2 \sqrt{2} \pi \alpha} \; m_{T}^{2} \;\; .
\eeq
The Lagrangian (\ref{G}) provides therefore a comprehensive description of
the electromagnetic properties of neutrinos propagating inside a plasma,
valid for the entire spectrum of both longitudinal and transverse modes. It
reduces to an electric charge when the hard regime applies, {\it i.e.}, in the
asymptotic limit of the transverse branch where plasmons behave as a
classical field.

\section{The plasmon decay rate}

Inside stars, plasma waves may decay into neutrino pairs as a result of the
effective coupling (\ref{G}). At the helium cores of red-giants where the
stellar density is fairly large, the plasma frequency reaches up to
$\omega_{0} \sim$ 20 keV. Once produced, the neutrinos do not interact
with stellar interiors. Most of the stars are indeed completely transparent to
these species which, therefore, escape freely from their cores. The
corresponding cooling rate depends on the distribution of plasma waves as
well as on the efficiency with which a plasmon is transmuted into a
neutrino pair.

The decay rate of plasmons obtains from the phase space integral
\beq
\Gamma_{D} \; = \; \frac{1}{2 \omega} \;
{\displaystyle \int} \;
\tilde{dk}_{\nu} \; \tilde{dk}_{\bar{\nu}} \;
(2 \pi)^{4} \, \delta^{4} \left( Q - k_{\nu} - k_{\bar{\nu}} \right) \;
{\displaystyle \sum}_{\rm \; spins} \;
\vert {\cal M} \left( \gamma_{Q} \to \nu \bar{\nu} \right) \vert^{2}
\;\; ,
\label{DECAYRATEA}
\eeq
where the relevant matrix element is
\beq
{\cal M} \left( \gamma_{Q} \to \nu \bar{\nu} \right) \; = \;
\frac{\sqrt{2} G_{F} g_{V}}{e} \; Q^{2} \;
\left\{ \bar{U}(k_{\nu}) {\rlap/{\epsilon}} L V(k_{\bar{\nu}}) \right\}
\;\; .
\eeq
In the Lorentz gauge, the polarization of transverse plasmons are the pure
vectors $\vec{\epsilon_{T}}$, transverse to the direction $\vec{q \,}$ of
motion, whilst the longitudinal polarization is taken care of by the
four-vector
\beq
\epsilon_{L}^{\mu} \; = \; (q , \frac{\vec{q \,}}{q} \omega_{L}) /
\sqrt{ \omega_{L}^{2} - q^{2} } \;\; .
\eeq
A straightforward calculation leads to the decay rate
\beq
\Gamma_{D} \; = \; \frac{1}{48 \pi^{2}} \;
\frac{G_{F}^{2} g_{V}^{2}}{\alpha} \; \frac{1}{\omega} \;
\left( Q^{2} \right)^{3} \;\; .
\label{DECAYRATEB}
\eeq
In the hard regime, where the electromagnetic properties of neutrinos are
well described by the electric charge $e Q_{\nu}$ -- see relations (\ref{E})
and (\ref{QNU}) -- the decay rate becomes
\beq
\Gamma_{D} \; = \; \frac{\alpha}{6} \; Q_{\nu}^{2} \;
\frac{Q^{2}}{\omega} \;\; .
\label{DECAYRATEC}
\eeq
The neutrino energy loss may be derived from the sum over the
Bose-Einstein distribution $\eta_{B}(\omega)$ of the individual decays
\beq
\epsilon_{Loss} \; = \; {\displaystyle \int} \,
\frac{d^{3} \vec{q \,}}{8 \pi^{3}} \; g_{h} \;
\eta_{B} (\omega) \;\ \omega \;
Z \; \Gamma_{D} \;\; .
\label{ENERGYLOSS}
\eeq
Each neutrino pair carries outside the star the energy $\omega$ of its
progenitor. The helicity factor $g_{h}$ respectively takes the values 1 and 2
for the longitudinal and transverse modes. The additional factor $Z$
corrects for the term $1 / 2 \omega$ in relation (\ref{DECAYRATEA}).
For ordinary particles with mass $m$, the relation between the energy $E$
and the momentum  $\vec{p \,}$ is merely
\beq
E^{2} \, - \, \vec{p \,}^{2} \; = \; P^{2} \; = \; m^{2} \;\; ,
\label{NORMAL}
\eeq
so that, when the Lorentz invariant expression
$d^{4} P \; \delta \left( P^{2} - m^{2} \right)$ is integrated over the
positive
values of $P^{0}$, it translates into the conventional phase space element
$d^{3} \vec{p \,} / \, 2 E$. For plasmons, expression (\ref{NORMAL}) is
replaced by the more intricate dispersion relations (\ref{DISPERSION})
which, once integrated over the positive values of $Q^{0}$, become
\beq
{\displaystyle \int} d^{4}Q \;\; \theta(Q^{0}) \;\;
\delta \left\{ Q^{2} \, + \, \Pi \left(Q^{0}, \vec{q \,}\right) \right\} \; =
\;
Z \; \frac{d^{3} \vec{q \,}}{2 \omega} \;\; ,
\eeq
with
\beq
Z^{-1} \; = \; \left| 1 \, + \,
\frac{\partial \Pi}{\partial \omega^{2}} \right| \;\; .
\eeq
For the transverse mode of a degenerate plasma, this normalisation factor
may be expressed as
\beq
Z_{T} \; = \; \frac{\displaystyle 2 \OMT^{2}
\left( \OMT^{2} \, - \, v_{F}^{2} q^{2} \right)}
{\displaystyle 3 \, \OMP^{2} \OMT^{2} \; + \;
\left( \OMT^{2} + q^{2} \right) \left( \OMT^{2} - v_{F}^{2} q^{2} \right)
\; - \; 2 \OMT^{2} \left( \OMT^{2} \, - \, q^{2} \right)} \;\; ,
\eeq
whereas, for the longitudinal branch, it becomes
\beq
Z_{L} \; = \; \frac{2 \OML^{2}}{Q^{2}} \; \left\{
\frac{\displaystyle \left( \OML^{2} \, - \, v_{F}^{2} q^{2} \right)}
{\displaystyle 3 \, \OMP^{2} \; - \;
\left( \OML^{2} \, - \, v_{F}^{2} q^{2} \right)} \right\} \;\; .
\eeq
In the non-relativistic regime for which the velocity $v_{F} \to 0$, $Z_{T}$
merely reduces to unity whereas $Z_{L}$ tends towards $\OMP^{2} / Q^{2}$.

Once again, similar expressions for the energy loss may be obtained in the
thermal field theory approach. The neutrino (antineutrino) production rate,
due for instance to transverse photon decay, is easily derived by using the
cutting rules of Kobes and Semenoff \cite{CR}, given by
\bea
R_{T} & = & \frac {dN_{\nu}}{d^4x} \nonumber \\
& = & \frac{2 G_{F}^{2} g_{V}^{2}}{e^{2}} \,
{\displaystyle \int} \, \frac {d^4Q} {(2\pi)^4}
{\displaystyle \int} \, \frac {d^4K} {(2\pi)^4} \;
2 \pi \delta(K^2) \; 2 \pi \delta\left[ (Q-K)^2-m^2 \right]
\left\{ \eta_{B }(\omega) \, + \, \theta(-Q^0) \right\} \nonumber \\
& & \times \,
2 \pi \, \delta \left[ Q^2-\mbox{Re} \, \Pi_{T}(Q) \right] \,
\mbox{Tr} \left[ {\rlap/K} \gamma^{\mu} ({\rlap/K} + {\rlap/Q})
\gamma^{\nu} \right] \,
{\cal P}^{\mu \nu}_T \; \left( Q^{2} \right)^{2} \;\; .
\ena
Here $K$ and $Q$ are respectively the neutrino and photon four-momenta.
The transverse photon projection operator is
\beq
{\cal P}^{ij}_T \, = \, - \,
\delta^{ij} \, + \, q^{i} q^{j}/q^2 \;\; ,
\eeq
with all its other components set equal to zero \cite{Wel}. After performing
the integrations, the result takes the final form
\beq
R_{T} \; = \; \frac{G_{F}^{2} g_{V}^{2}}{48 \pi^{4} \alpha} \;
{\displaystyle \int_{0}^{\infty}} \; \frac{q^{2} dq}{\omega} \;
Z_{T} \; \eta_{B}(\omega) \; \left( Q^2 \right)^{3} \;\; .
\eeq
The energy loss rate may be inferred directly from the previous integral~:
\beq
\epsilon_{T} \; = \; \frac{G_{F}^{2} F_{\nu}}{48 \pi^{4} \alpha} \;
{\displaystyle \int_{0}^{\infty}} \; q^{2} dq \;
Z_{T} \; \eta_{B}(\omega) \; \left( Q^2 \right)^{3} \;\; ,
\label{LOSSTEXACT}
\eeq
where the factor $F_{\nu}$ takes into account the three neutrino species
produced in plasmon decay
\beq
F_{\nu} \; = \; {\displaystyle \sum}_{\nu} \; g_{V}^{2}(\nu) \; = \;
{\displaystyle \frac{3}{4}} - 2 \, \sin^{2} \theta_{W} + 12 \, \sin^{4}
\theta_{W} \;\; .
\eeq
As regards the longitudinal branch, a similar expression obtains where the
plasmon momentum is integrated up to $q_{\, max}$
\beq
\epsilon_{L} \; = \; \frac{G_{F}^{2} F_{\nu}}{96 \pi^{4} \alpha} \;
{\displaystyle \int_{0}^{q_{\, max}}} \; q^{2} dq \;
Z_{L} \; \eta_{B}(\omega) \; \left( Q^2 \right)^{3} \;\; .
\label{LOSSLEXACT}
\eeq
In the next section, the behaviour of the exact cooling rate $\epsilon =
\epsilon_{T} + \epsilon_{L}$ will be discussed in various stellar
environments. Two approximations of this exact energy loss may be
defined~:

(i) The estimate $\epsilon^{Q}$ obtains when the charge radius
(\ref{CRC}) is replaced by the electric charge (\ref{E}), so that the
transverse
cooling is now
\beq
\epsilon^{Q}_{T} \; = \; \frac{16 \alpha G_{F}^{2} F_{\nu}}{3 \pi^{2}} \;
{\cal I}_{P}^{2} \;
{\displaystyle \int_{0}^{\infty}} \; q^{2} dq \;
Z_{T} \; \eta_{B}(\omega) \; Q^2 \;\; ,
\label{LOSSTCHARGE}
\eeq
whereas the longitudinal energy loss is
\beq
\epsilon^{Q}_{L} \; = \; \frac{8 \alpha G_{F}^{2} F_{\nu}}{3 \pi^{2}} \;
{\cal I}_{P}^{2} \;
{\displaystyle \int_{0}^{q_{\, max}}} \; q^{2} dq \;
Z_{L} \; \eta_{B}(\omega) \; Q^2 \;\; .
\label{LOSSLCHARGE}
\eeq

(ii) The non-relativistic approximation corresponds to the limit
where $v_{F}$ vanishes. The plasma frequency is therefore defined as the
ratio
\beq
\omega_{0}^{2} \; = \; \frac{4 \alpha}{3 \pi} \; \frac{p_{F}^{3}}{m} \;\; .
\label{OMEGAPLASMANR}
\eeq
As the dispersion relations (\ref{LONGITUDINALNR}) and
(\ref{TRANSVERSENR}) now apply, the expressions for the energy loss
rate $\epsilon^{NR}$ considerably simplify. We simply recover
the results of Adams, Ruderman and Woo \cite{ARW}, later on corrected
by Zaidi \cite{Plasmon}. Neutrino emission by transverse photons occurs at
a
pace~:
\beq
\epsilon^{NR}_{T} \; = \;
\frac{G_{F}^{2} F_{\nu}}{48 \pi^{2}\alpha} \; \OMP^{6} \; n_{\gamma}
\;\; ,
\label{LOSSRATENRT}
\eeq
where the photon density $n_{\gamma}$ is defined by
\beq
n_{\gamma} \; = \; \frac{1}{\pi^{2}} \,
{\displaystyle \int}_{0}^{+ \infty} \, q^{2} dq \; \eta_{B}(\OMT) \;\; .
\eeq
As regards the energy dragged away from the longitudinal mode, the
corresponding loss rate may be expressed exactly
\beq
\epsilon^{NR}_{L} \; = \;
\frac{G_{F}^{2} F_{\nu}}{1260 \, \pi^{4}\alpha} \; \OMP^{9} \;
\eta_{B}(\OMP) \;\; .
\label{LOSSRATENRL}
\eeq

\section{Discussion}

Neutrino cooling dominates the slow contraction of red-giant stars
\cite{Raf}, whose core temperature steadily increases in time, up to the
point where, at the tip of the red-giant branch of the Hertzsprung-Russell
diagram, the triple-alpha reaction sets on. For low-mass stars, typical
conditions at the center are $\rho \sim 10^{6}$ g cm$^{-3}$ with $T \simeq
10^{8}$ K. During this stage of their evolution, stars have a burning shell of
hydrogen surrounding a degenerate helium core whose mass
increases at a pace dominated by the neutrino losses. Thus, the mass of
helium processed on the red-giant branch may be related to plasmon decay
and its associated cooling \cite{RGCOOLING}. As a matter of fact,
observations of the color-magnitude diagrams of globular clusters offer an
interesting handle on the
cooling of red-giant cores by neutrino emission. The latter mechanism also
dominates the early evolution of white-dwarves, slightly after their
formation, when they are still relatively hot, with central temperature $T
\sim 10^{7}$ to $10^{9}$ K. The cooling of white-dwarves is expected to be
measured in the near future, using asteroseismology techniques. The slow
variation of the period of pulsating white-dwarves may be directly related to
the decrease of their central temperature \cite{WDCOOLING}.

The case of red-giant stars is embraced in fig.~3a where the exact energy loss
rate $\epsilon$ varies with the density $\rho$, at fixed temperature $T =
10^{8}$ K and fixed electron fraction $Y_{e} = 0.5$ (pure helium). For small
values of the density $\rho$, the Fermi momentum
\beq
p_{F} \; = \; \left( 3 \pi^{2} Y_{e} \rho / m_{u} \right)^{1/3}
\eeq
is smaller than the electron mass $m$, and the non-relativistic regime
applies. The point $\rho = 10^{5}$ g cm$^{-3}$, for instance, corresponds to
$p_{F} = 190$ keV, so that the plasma frequency is $\OMP \simeq 6$ keV.
Even at such a low density, electrons are degenerate because the
temperature $T \simeq 9$ keV is negligible with respect to the Fermi
momentum $p_{F}$. Since $T$ exceeds the plasma frequency $\OMP$ in
the left-hand side region of fig.~3a, relations (\ref{LOSSRATENRT}) and
(\ref{LOSSRATENRL}) simplify respectively into
\beq
\epsilon^{NR}_{T}(T > \OMP) \; = \;
\frac{\zeta (3)}{24 \, \pi^{4}\alpha} \; G_{F}^{2} F_{\nu} \; \OMP^{6} \;
T^{3}
\;\; ,
\eeq
and
\beq
\epsilon^{NR}_{L}(T > \OMP) \; = \;
\frac{G_{F}^{2} F_{\nu}}{1260 \, \pi^{4}\alpha} \; \OMP^{8} \; T \;\; .
\eeq
The transverse rate overshadows completely the longitudinal emission.
Note that at low density, the energy loss $\epsilon \simeq
\epsilon^{NR}_{T} + \epsilon^{NR}_{L}$ increases with $\rho$. On
the contrary, in the high density regime (right-hand side domain of fig.~3a),
electrons are ultra-relativistic and the temperature is now much smaller
than the plasma frequency, hence a decrease of $\epsilon$ with $\rho$. The
Boltzmann factor $\exp (- \OMP / T)$ drops as there are fewer and fewer
plasmons. The maximum which the curve exhibits corresponds to the
transition between these two regimes.

In fig.~4a, devoted to white-dwarf cooling, the density has been set equal
to the typical value of $1.8$ tons cm$^{- 3}$, a pure carbon composition has
been assumed ($Y_{e} = 0.5$), and the temperature is varied from $10^{6}$
up to $10^{9}$ K. The Fermi momentum $p_{F} \sim 500$ keV corresponds
to the relativistic velocity $v_{F} = 0.7$ and to the plasma frequency $\OMP
= 23.2$ keV. Note also that the transverse mass of high-energy photons is
$m_{T} = 24.7$ keV. Since the maximal temperature considered here, $T =
86$ keV, is much smaller than the Fermi momentum, the gas is still
degenerate. The increase of the energy loss rate $\epsilon$ with $T$ is
impressive.

The conditions of density and temperature in the plots~3b and 4b are
respectively the same as in fig.~3a and 4a. The solid line stands for the ratio
$\epsilon^{Q} / \epsilon$ and corresponds to the electric charge
approximation. The short dash curve refers to the non-relativistic estimate
$\epsilon^{NR} / \epsilon$. The numerical fit published by Itoh {\it et al.}
\cite{IMHK} is extensively used for astrophysical purposes. Its ratio to the
exact cooling rate  is represented here by the dotted curve while the short
dash-dotted line stands for the recent numerical estimate by Haft {\it et
al.} \cite{HRW}. The energy loss rate $\epsilon$ is a function of both the
density $\rho$ and the temperature $T$. In a degenerate medium, which is
the situation under scrutiny here, the density alone determines the Fermi
momentum, the plasma frequency as well as the dispersion relations of the
plasmon propagation. Therefore, the structure of both transverse and
longitudinal modes depends only on the parameter $\rho$. As for the
temperature, it conditions the filling of the plasma states along the various
branches and governs, for instance, the above-mentionned density
$n_{\gamma}$. Some remarks are in order.

(A) In the regime where the temperature exceeds the plasma frequency, the
high-energy tail of both transverse and longitudinal branches are filled up.
Note that $\OML^{2} - q^{2}$ vanishes near the tip of the longitudinal
branch. Therefore, the decays of longitudinal plasmons are suppressed and
neutrino cooling is dominated by the behaviour of the high-energy, {\it
i.e.}, hard, transverse photons. The loss rate $\epsilon^{Q}$ is a good
estimate of $\epsilon$ in the high temperature regime. That is why the
ratio $\epsilon^{Q} / \epsilon$ (solid curve) becomes unity in the left-hand
side of fig.~3b and in the right-hand portion of fig.~4b.

(B) At low temperature, {\it i.e.}, in the limit where $\OMP$ exceeds $T$,
the low-energy plasma states only are thermally excited, and the average
momentum $q$ is, a priori, much smaller than the plasma frequency.
Plasmons may be pictured as if they were at rest, with energy $\OMP$,
vanishing momentum $q$ and therefore an effective mass of order the
plasma frequency $\OMP$. The exact decay rate (\ref{DECAYRATEB})
varies as $(Q^{2})^{3}$ whereas the approximate relation
(\ref{DECAYRATEC}) is proportional to $m_{T}^{4} \, Q^{2}$. Therefore,
when $T$ is small with respect to $\OMP$, the ratio $\epsilon^{Q} /
\epsilon$
behaves roughly as $(m_{T} / \OMP)^{4}$. In fig.~3b for instance, the
solid curve tends asymptotically towards the ultra-relativistic limit of
$(3/2)^{2} = 2.25$ when $\rho$ is large.
In the low temperature part of fig.~4b, it converges
towards the value $(24.7 / 23.2)^{4} \simeq 1.3$ (see above). There is
nevertheless an exception to this behaviour. When the plasma is
non-relativistic, the longitudinal branch is noticeably flat and, unlike for
the
other mode, all its propagation states are equally filled up, even at extremely
low temperature. The momentum $q$ is not forced to vanish because the
Bose-Einstein factor $\eta_{B}(\OMP)$ is constant throughout the
longitudinal branch. In this regime, the energy loss
\beq
\epsilon^{NR}_{L}(T < \OMP) \; = \;
\frac{G_{F}^{2} F_{\nu}}{1260 \, \pi^{4}\alpha} \; \OMP^{9} \;
e^{\displaystyle - \OMP / T}
\eeq
eclipses completely the transverse production of neutrinos
\beq
\epsilon^{NR}_{T}(T < \OMP) \; = \;
\frac{G_{F}^{2} F_{\nu}}{48 \sqrt{2} \, \pi^{7/2} \, \alpha} \;
\OMP^{15/2} \; T^{3/2} \; e^{\displaystyle - \OMP / T} \;\; .
\eeq
Therefore, at low temperature and for a non-relativistic degenerate electron
gas, the electric charge approximation $\epsilon^{Q}$ overestimates the
exact rate by a factor of $35 / 8 \simeq 4.4$.

(C) At low density, the non-relativistic regime obtains and $\epsilon^{NR}$
is a good estimate of the exact loss rate. Consequently, in the left-hand side
of fig.~3b, the short dash curve approaches unity. On the contrary, for large
values of the density, the non-relativistic expression
(\ref{OMEGAPLASMANR}) overestimates slightly the plasma frequency
$\OMP$. In this regime, the loss rate $\epsilon$ behaves as $\exp (- \OMP
/ T)$, hence the exponential drop which the ratio $\epsilon^{NR} /
\epsilon$ exhibits as the density increases (right-hand side of fig.~3b) or as
the temperature diminishes (left-hand side of fig.~4b).

(D) For red-giant stars, the fit which Haft {\it et al.} propose is
impressively
good.
Notice how the short dash-dotted curve remarkably sticks to unity
throughout fig.~3b. The Japanese approximation slightly underestimates
the exact rate below $\rho = 10^{6}$ g cm$^{-3}$. As regards white-dwarf
cooling, Haft's fit is still fairly good, especially between $10^{7}$ and
$10^{9}$ K. On the contrary, the result by Itoh {\it et al.} exhibits wiggles
at
the transition temperatures of the numerical estimate whose quality is not
completely satisfactory. Furthermore, below $T = 10^{7}$ K, the
corresponding dotted curve diverges whereas Haft's fit still gives an
accurate prediction.
\bigskip

The coherent diffusion of neutrinos on the electrons of a plasma induces an
effective interaction with the electromagnetic field. An intuitive approach
as well as a more formal discussion performed in the framework of thermal
field theory have been presented here. Both methods lead to the same
conclusions. Neutrinos which propagate inside matter acquire an effective
charge radius of order $6.5 \times 10^{-16}$ cm $g_{V}^{1/2}$, regardless of
the specific properties of the plasma. The description of the electromagnetic
interactions of neutrinos by a charge radius, strictly valid in the limit of
the
soft-loop regime, turns out to be more general and may apply to the entire
spectrum of plasmons. In particular, in the hard limit of high-energy
transverse quanta, this neutrino charge radius translates into a mere electric
charge. Since plasmon decay is an efficient mechanism of stellar energy loss
when the cores of stars are hot, {\it i.e.}, in a regime where the estimate
$\epsilon^{Q}$ is accurate, the thermal electric charge $Q_{\nu}$ of
neutrinos is therefore relevant to astrophysics.
However minute as it may look, it is
nevertheless responsible for the evolution of red-giant stars as well as for
the early cooling of nascent white-dwarves.

Suppose finally that the tau neutrino has a large charge radius of order $6.5
\times 10^{-16}$ cm. Plasmon decay into $\nu_{\tau}$ pairs would induce
an extra neutrino emission comparable to the standard production.
Observations of the magnitude shift between the tip of the red-giant branch
and the horizontal sequence of the HR diagram preclude the existence of
any additional neutrino cooling with a magnitude larger than the standard
energy loss. We readily infer a limit of order $6 - 7 \times 10^{-16}$ cm on
any anomalous charge radius. For electron and muon neutrinos, the
laboratory constraints, of order $1 - 3 \times 10^{-16}$ cm, are already more
stringent \cite{BEAMDUMP}. However, for tau neutrinos, the stellar limit
compares fairly well to the upper bound of $8 - 9 \times 10^{-16}$ cm
derived from the ASP and CELLO experiments \cite{ASP}.

\newpage
\noindent{\large\bf Figure Caption}
\vskip 1.cm
\begin{description}
\item[Fig.~1] The neutrino current diffuses coherently on the electrons (1a
and 1b), and on the positrons (1c and 1d) of the thermal bath. Through its
interactions with the spectator particles, this current couples to the
electromagnetic field $A^{\beta}$, hence an effective electric charge (or
charge radius) for the neutrinos which propagate inside a plasma.

\item[Fig.~2] One-loop Feynman diagram contribution to the
electromagnetic interactions of neutrinos, in finite temperature field theory.

\item[Fig.~3] The total neutrino emissivity $\epsilon$ as a function of the
plasma density $\rho$ for pure helium, at $T=10^8$ K. Fig.~3a shows the
exact absolute rate whilst fig.~3b presents the ratios between various
approximations and the exact rate.

\item[Fig.~4] The neutrino emissivity as a function of the plasma
temperature for pure carbon, at $\rho=1.8 \times 10^6$ g/cm$^3$. Fig.~4a
shows the exact absolute rate whilst fig.~4b exhibits the ratios between
various approximations and the exact rate.
\end{description}


\newpage
\begin{thebibliography}{99}
\bibitem{ARW} J.~B.~Adams, M.~A.~Ruderman and C.~H.~Woo, {\sl
Phys. Rev.}~{\bf 129} (1963) 1383.

\bibitem{Plasmon} M.~H.~Zaidi, {\sl Nuovo Cim.}~{\bf 40A} (1965) 502;
G.~Beaudet, V.~Petrosian and E.~E.~Salpeter, {\sl Ap. J.}~{\bf 150} (1967)
979; D.~A.~Dicus, {\sl Phys. Rev.}~{\bf D6} (1972) 941.

\bibitem{Bra} E.~Braaten, {\sl Phys. Rev. Lett.}~{\bf 66} (1991) 1655.

\bibitem{IMHK} N.~Itoh, H.~Mutoh, A.~Hikita and Y.~Kohyama, {\sl Ap.
J. }~{\bf 395} (1992) 622. See also H.~Munakata, Y.~Kohyama and N.~Itoh,
{\sl Ap. J.}~{\bf 296} (1985) 197.


\bibitem{BS} E.~Braaten and D.~Segel, {\sl Phys. Rev.}~{\bf D47} (1993)
1478.

\bibitem{HRW} M.~Haft, G.~Raffelt and A.~Weiss, to appear in {\sl Ap. J.}

\bibitem{RN} D.~Notzold and G.~Raffelt, {\sl Nucl. Phys.}~{\bf B307} (1988)
924; P.~B.~Pal and T.~N.~Pham, {\sl Phys. Rev.}~{\bf D40} (1989) 259.

\bibitem{Charge} A.~Grau and J.~A.~Grifols, {\sl Phys. Lett.}~{\bf B166}
(1986) 233; V.~N.~Oraevskii and V.~B.~Semikoz, {\sl Physica}~{\bf 142A}
(1987) 135; V.~N.~Oraevskii and V.~N.~Ursov, {\sl Phys. Lett.}~{\bf B209}
(1988) 83; J.~C.~D'Ollivo, J.~F.~Nieves and P.~B.~Pal, {\sl Phys. Rev.}~{\bf
D40} (1989) 3679; T.~Altherr and K.~Kainulainen, {\sl Phys. Lett.}~{\bf B262}
(1991) 79; J.~F.~Nieves and P.~B.~Pal, Univ. of Texas preprint, CPP-93-12.

\bibitem{LvW} N. P. Landsman and Ch. G. van Weert, {\sl Phys. Rep.}~{\bf
145} (1987) 141.

\bibitem{HARDSOFT} E.~Braaten and R.~Pisarski, {\sl Nucl. Phys.}~{\bf
B337} (1990) 569; {\sl Nucl. Phys.}~{\bf B339} (1990) 310.

 \bibitem{Wel} H.~A.~Weldon, {\sl Phys. Rev.}~{\bf D26} (1982) 1394.

\bibitem{APR} T.~Altherr, E.~Petitgirard and T.~del Rio Gaztelurrutia,
{\sl Astropart. Phys.}~{\bf 1} (1993) 289.

\bibitem{Jan} B.~Jancovici, {\sl Nuovo Cim.}~{\bf 25} (1962) 428.

\bibitem{CR} H.~A.~Weldon, {\sl Phys. Rev.}~{\bf D28} (1983) 2007;
R.~L.~Kobes and G.~W.~Semenoff, {\sl Nucl. Phys.}~{\bf B260} (1985) 714
and {\bf B272} (1986) 329; N.~Ashida, H.~Nakkagawa, A.~Ni\'egawa and
H.~Yokota, {\sl Ann. Phys. (NY)}~{\bf 215} (1992) 315.

\bibitem{Raf} G.~Raffelt, {\sl Ap. J.}~{\bf 365} (1990) 559.

\bibitem{RGCOOLING} A.~V.~Sweigart and P.~G.~Gross, {\sl
Astrophys. J. Suppl.}~{\bf 36} (1978) 405.

\bibitem{WDCOOLING} S.~O.~Kepler {\it et al.}, {\sl Ap. J.}~{\bf
378} (1991) L45. See also the analysis by J.~Isern, M.~Hernanz and
E.~Garcia-Berro, {\sl Ap. J.}~{\bf 392} (1992) L23.

\bibitem{BEAMDUMP} K.~Abe {\it et al.}, {\sl Phys. Rev. Lett.}~{\bf 58}
(1987) 636; J.~Dorenbosch {\it et al.}, {\sl Zeit. fur Phys.}~{\bf C41}
(1989) 567; L.~A.~Ahrens {\it et al.}, {\sl Phys. Rev.}~{\bf D41} (1990) 3297;
I.~I.~Gurevich {\it et al.}, Kurchatov Institute of Atomic Energy preprint
(1993).

\bibitem{ASP} CELLO Collaboration, H.~J.~Behrend {\it et al.}, {\sl Phys.
Lett.}~{\bf B215} (1988) 186; C.~Hearty {\it et al.}, {\sl Phys. Rev.}~{\bf
D39} (1989) 3207.
\end{thebibliography}
\end{document}